# First Detection of the Powerful Gamma Ray Burst GRB221009A by the THEMIS ESA and SST particle detectors on October 9, 2022.

O.V. Agapitov[1,2], M. Balikhin[3], A. J. Hull[2], Y. Hobara[4,5,6], V. Angelopoulos[7], F.S. Mozer[2]

**Abstract**

We present the first results study of the effects of the powerful Gamma Ray Burst GRB 221009A that occurred on October 9, 2022, and was serendipitously recorded by electron and proton detectors aboard the four spacecraft of the NASA THEMIS mission. Long-duration gamma-ray bursts (GRBs) are powerful cosmic explosions, signaling the death of massive stars, and, among them, GRB 221009A is so far the brightest burst ever observed due to its enormous energy ($E_{\gamma\ iso} \approx 10^{55}$ erg) and proximity (the redshift is z≈0.1505). The THEMIS mission launched in 2008 was designed to study the plasma processes in the Earth's magnetosphere and the solar wind. The particle flux measurements from the two inner magnetosphere THEMIS probes THA and THE and two outer probes (renamed ARTEMIS after 2010) THB and THC orbiting the Moon captured the dynamics of GRB 221009A with a high-time resolution of 4 (up to 8) measurements per second. This allowed us to resolve the fine structure of the gamma-ray burst and determine the temporal scales of the two main bursts' spiky structure complementing the results from gamma-ray space telescopes and detectors.

**Introduction**

GRB 221009A was a bright and long-lasting gamma-ray burst (GRB) detected by the Burst Alert Telescope (BAT) aboard the Swift satellite, an hour later by the Gamma-ray Burst Monitor (GBM) (Dichiara et al. 2022) on board the Fermi Gamma-ray Space Telescope (FGST) (Veres et al., 2022; Bissardi et al. 2022; Lesage et al. 2022; Pilera et al. 2022), by other space observatories such as Swift (Dichiara et al. 2022), AGILE (Piano et al. 2022; Ursi et al. 2022), INTEGRAL (Gotz et al. 2022), Solar Orbiter (Xiao et al. 2022), SRG (Lapshov et al. 2022), Konus (Frederiks et al. 2022, 2023), GRBAlpha (Ripa et al. 2022), and STPSat-6 (Mitchell et al. 2022), High Energy Burst Searcher (HEBS) on SATech-01 (Liu et al. 2022), by the ground observatory LHAASO (The Large High Altitude Air Shower Observatory) with striking very-high energy features (Huang et al. 2022). The GBM light curve consists of an initial ~10 s long pulse at 13:16:59 UTC, followed by an extraordinarily bright episode roughly ~180 s after the trigger time, lasting at least 100 seconds (Veres et al., 2022). The afterglow outburst outshone all other GRBs seen before (Sahu et al., 2023). The Large High Altitude Air Shower Observatory (LHAASO) with the water Cherenkov detector array (WCDA) and the larger air shower kilometer square area (KM2A) detector observed more than 5000 very high energy (VHE) photons in the 500 GeV–18 TeV energy range within 2000 s from the trigger, making them the most energetic photons ever observed from a GRB (Huang et al. 2022). The event was so long and intense that it caused sudden Earth global ionospheric disturbances (both day and night) - a result of the increased ionization by X- and γ-ray emission (Hayes and Gallagher, 2022; Pal et al., 2023) from the VLF/LF sub-ionospheric signals dynamics in the D-region of Earth's ionosphere (∼60–100 km). The optical observations of this burst (located at around RA = 288.282 and Dec = 19.495 (Pillera et al. 2022)) have a relatively small redshift z = 0.1505 (Castro-Tirado et al. 2022; de Ugarte Postigo & Izzo 2022) compared to most other long bursts, which indicates that this is one of the closest observed long-duration GRB (GRB 211211A had even lower red shift of 0.076, which corresponded to a distance of ~346 megaparsecs (Rastinejad rt al., 2022; Troja et al., 2022; Mei et al., 2022; Gompertz et al. 2022). The total emitted isotropic-equivalent gamma-ray energy from GRB 221009A is estimated to be $(2–6) \times 10^{54}$ erg (de Ugarte Postigo et al. 2022; Kann & Agui 2022).

---
[1] Corresponding author agapitov@ssl.berkeley.edu
[2] Space Sciences Laboratory, University of California, Berkeley, CA 94720
[3] University of Sheffield, Sheffield, UK
[4] Graduate School of Informatics and Engineering, The University of Electro-Communications (UEC), Chofu, Tokyo 1828585, Japan
[5] Center for Space Science and Radio Engineering, UEC, Chofu, Tokyo 1828585, Japan
[6] Research Center for Realizing Sustainable Societies, UEC, 1-5-1 Chofugaoka, Chofu, Tokyo 1828585, Japan
[7] University of California Los Angeles, Los Angeles, CA

The high energy gamma photons guide the sputtering of secondary particles (proton, electrons, and probably secondary photons) from the material of the spacecraft, which can be detected by particles detectors on board spacecraft (Pisacane 2008). Similar effects were reported by Schwartz et al. (2005) and Terasawa et al. (2005) after the gamma-ray giant flare of SGR 1806220. These measurements from spacecraft particle detectors provide an alternative perspective that complements observations from the gamma ray telescopes by providing high sampling rates capable of resolving the fine structure of bursts (Schwartz et al. 2005; Terasawa et al. 2005). GRB 221009A was detected in the electron flux measurements made by the HEPP-L charged particle detector on board the low Earth orbit (LEO) China Seismo-Electromagnetic Satellite (Battiston et al. 2023). Battiston et al. (2023) showed that the recorded anomalous signal of GRB221009A in the electron fluxes originated from secondary electrons produced via photon absorption in the passive material of the detector. The signal dynamics followed quite well the structure of the GRB221009A signal recorded by HEBS (Liu et al. 2022). Here we report observations of the fine structure of the GRB 221009A gamma burst signal recorded in electron and proton flux measurements (the bursts of secondary electrons produced through interactions of gamma photons with material of the detectors (Battiston et al. 2023)) by the electrostatic analyzers ESA (McFadden et al., 2008) and solid state telescopes SST (Larson et al., 2010) on board the four (of the five) THEMIS spacecraft (Angelopoulos et al., 2008) on October 9, 2022. The configuration of the spacecraft is shown in Figure 1. Specifically, we use data from THB and THC, orbiting the Moon as part of the ARTEMIS mission (Angelopoulos et al., 2015) and the two inner magnetosphere probes THA and THE. The THEMIS ESA and SST provide proton and electron distribution functions with sampling rate of the spin period (3 s for THA, THD, THE and 4 s for THB and THC). Battiston et al. (2023) showed that the signal from secondary electrons provide a full solid angle coverage, which allows us to decompose the spin resolution to individual measurements and increase the temporal resolution to 0.25 s (up to 0.125 s) to resolve the fine structure of the burst. The data recorded by SST and ESA detectors are unsaturated during the entire interval of observations, and resolves the fine structure of most intensive bursts, supplementing high-resolution measurements of the most intense period of the burst activity (some of gamma telescopes experienced saturation periods) made by HEBS (Liu et al. 2022, unsaturated data with subsecond timing resolution) and by HEPP-L (Battiston et al. 2023, unsaturated electron flux data with a second sampling rate).

**Data and Methods**

THEMIS (Time History of Events and Macroscale Interactions during Substorms) is NASA'a (National Aeronautics and Space Administration) mission, which consists of the five identically equipped satellites (probes THA, THB, THC, THD, and THE). The main goal of this mission is to carry out multipoint investigations of substorm phenomena in the tail of the terrestrial magnetosphere (Sibeck & Angelopoulos, 2008). After starting the Acceleration, Reconnection, Turbulence, and Electrodynamics of the Moon's

Interaction with the Sun (ARTEMIS) mission, which is a spin-off from THEMIS by repositioning two of the five THEMIS probes (THB and THC) in coordinated, lunar equatorial orbits, at distances of ∼55–65 geocentric earth radii, $R_E$ (∼1.1–12 selenocentric lunar radii, $R_L$). They perform systematic, two-point observations of the distant magnetotail, the solar wind, and the lunar space and planetary environment (Angelopoulos et al., 2010). The three inner probes (THA, THD, and THE) continue to collect data in the inner magnetosphere (∼1.2–9 $R_E$).

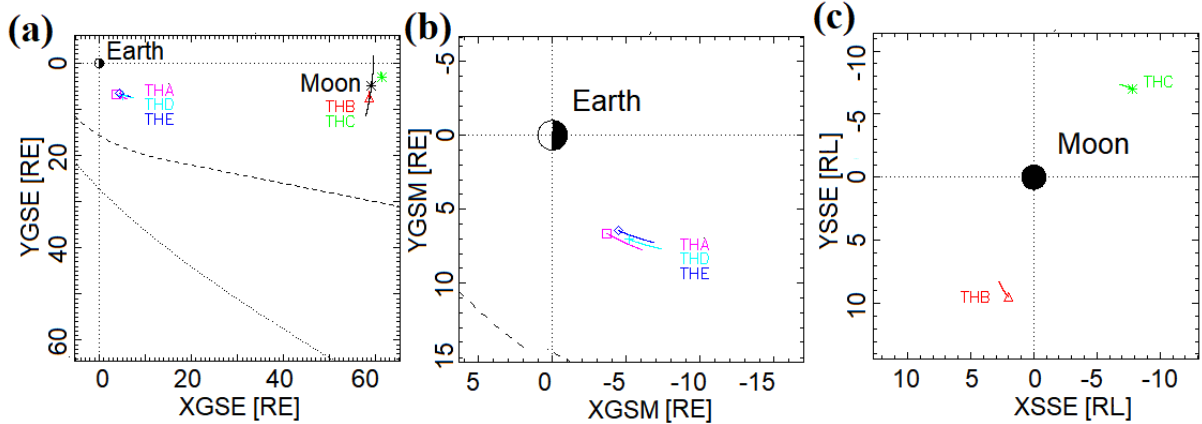

**Figure 1.** The configuration of THEMIS and ARTEMIS spacecraft during the GRB 211211A event is shown in panel (a) with the traces during the interval of 1200-1400 UT shown in the ecliptic plane. The modeled locations of Earth's magnetopause and bow shock are indicated by the dashed and dotted curves respectively. (b) – the inner THEMIS spacecraft configuration. (c) – the ARTEMIS spacecraft THB and THC in the selenocentric system (the ecliptic plane projection).

A pair of back-to-back top hat hemispherical electrostatic analyzers (ESA) measure the distribution functions of ions (0.005 to 25 keV) and electrons (0.005 to 30 keV) over 4π str to produce 3 s time resolution plasma moments (McFaddenetal.2008). The instrument consists of a pair of "top hat" electrostatic analyzers with common 180°×6° fields-of-view that sweep out 4π str each 3 s spin period. The sensors generally swept in energy (logarithmically) from ∼32 keV for electrons and ∼25 keV for ions, down to ∼6–7 eV. Nominal operations have 32 sweeps per spin, with 31 energy samples per sweep, plus one sample energy retrace, resulting in a typical measurement resolution of $\Delta E/E$ ∼32%. Particle events are registered by microchannel plate (MCP) detectors. At low-time-resolution, THEMIS generally maintain the "full" 32 energies sampled, and have an 88 bins composed solid-angle map. Combining the energy and solid-angle measurements can sufficiently improve time resolution. Solid-state telescopes (SST) measure the superthermal (0.02–1 MeV) part of the ion and electron distributions over 4π str with similar measurement regimes.

The configuration of THEMIS and ARTEMIS spacecraft during the GRB 221009A event is shown in Figure 1. Two THB and THC (ARTEMIS) were located near the Moon in the tail magnetosphere at geocentric distances of ∼58 $R_E$ (THB) and 61 $R_E$ (THC) from Earth. The inter-spacecraft distance was ∼3.5 RE (∼13 $R_L$). The inner probes were at geocentric distances of ∼8.5 $R_E$, with an inter-spacecraft distance of ∼1 $R_E$.

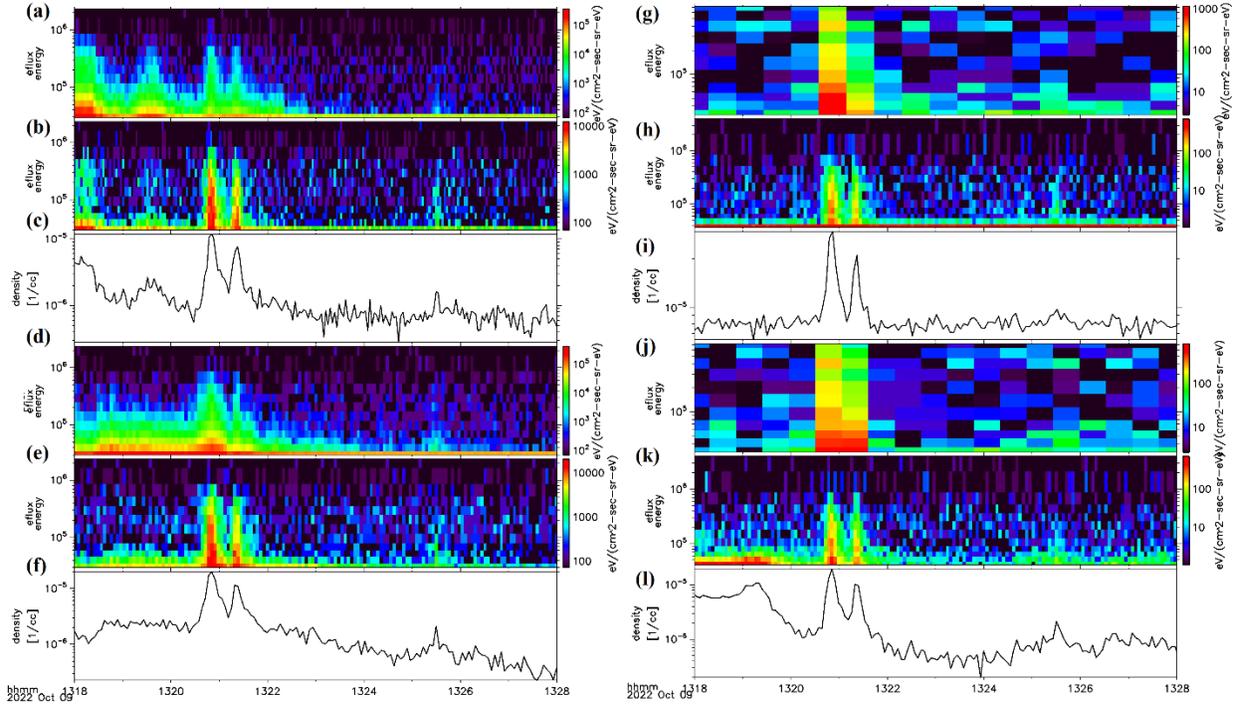

**Figure 2.** The electron and proton flux recorded by the SST detectors on board the THA and THE probes: (a) – the proton flux from the THA SST detector; (b) – the electron flux from the THA SST detector; (c) – the proton density in the SST energy range. Panels (d-f) present the same characteristics recorded on board THC, panels (g-i) - on board THB and panels (j-l) - on board THC.

The four spacecraft (THA, THB, THC, and THE) detected the effects of the gamma ray burst GRB221009A, which appear as two very intense bursts at 13:20:36-13:21:30 and a subsequent less intense burst at around 13:25:30 (Figure 2). The fifth THEMIS spacecraft, THD, did not collect particle data during this time interval. The ARTEMIS spacecraft THB and THC carried particles measurements in the same mode: spin resolution ESA and spin resolution SST (protons). The omnidirectional electron and proton flux dynamics recorded on board THA, THB, THC, and THE by the SST detectors are shown in Figure 2. The electron and proton apparent flux enhancements associated with GRB221009A are observed at energies up to 1 MeV in the SST measurements following the GRB221009A dynamics reported by Xiao et al. (2022) from Solar Orbiter STIX measurements of five energy bands in the range between 4 -150 keV (with the temporal resolution of 4 s) and by Liu et al. (2022) from HEBS detection in the range from 400-6000 keV. The decrease of the flux recorded by SST with energy above 1 MeV is in an agreement with the results of HEPP-L (Battiston et al. 2023). The data available from THA and THE was at spin averaged sampling rate of 3 s, so the processing of sub-second data could be applied only to the ARTEMIS probes THB and THC. The distance between the inner probes (THA, THD, and THE) and the ARTEMIS spacecraft (THB and THC) corresponded to ~1 light second (less than 0.25 seconds taking into account the geometry of the event). The spin resolution data from THA and THE do not sufficiently resolve the time delay of the GRB signal between the spacecraft. In the following we compare the timing of the signal recorded by THB and THC (available with the better time resolution) with the time of the signals from HEPP-L (Battiston et al. 2023) and HEBS (Liu et al. 2022) recorded on board the LEO spacecraft.

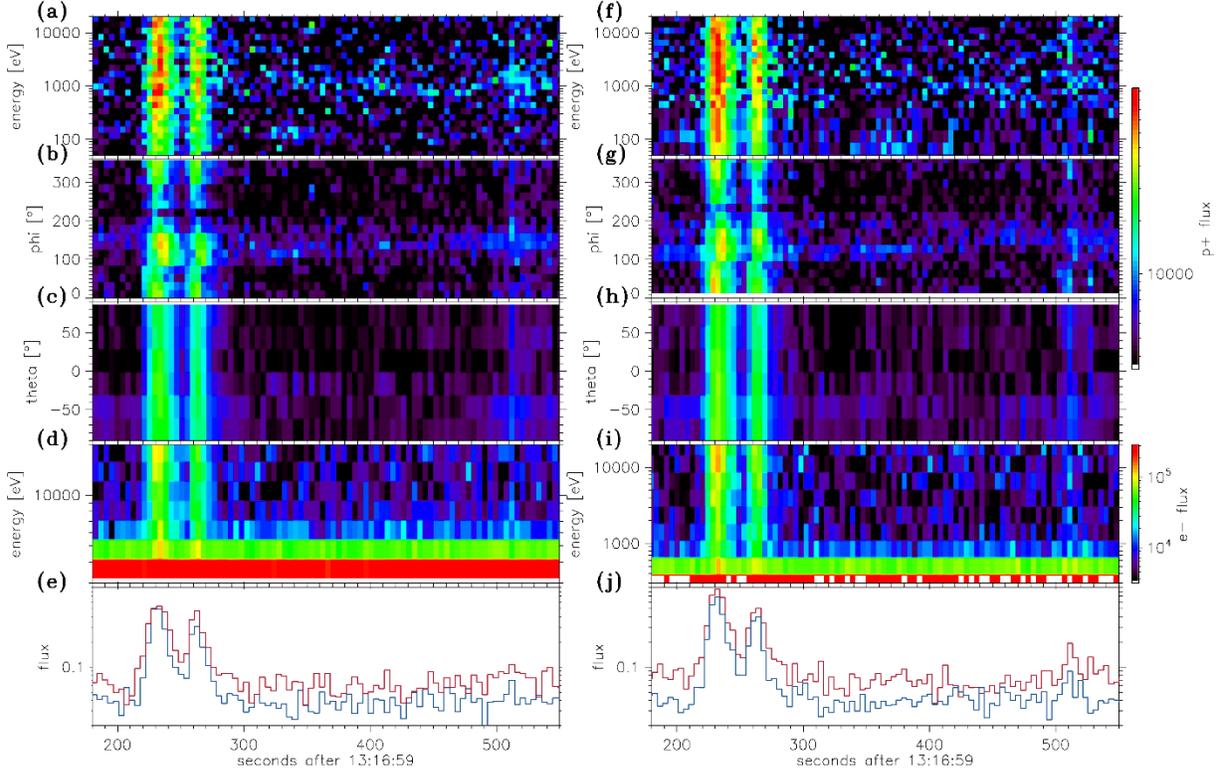

**Figure 3**. The apparent electron and proton fluxes recorded on board THB and THC probes with electrostatic analyzers (ESA): (a) – the omnidirectional proton flux from the THB ESA detector; (b) – the energy integrated proton flux on the azimuthal angle phi; (c) – the energy integrated proton flux on the polar angle theta; (d) – the omnidirectional electron flux from the THB SST detector; (e) – the integrated proton (red) and electron (blue) fluxes. Panels (f-j) present the same characteristics from THC.

The polar and azimuthal angular resolution of the ESA and SST detectors provide information on the directionality of particle fluxes. The THEMIS spacecraft THB and THC collected spin-resolution ESA proton and electron data shown in Figure 3. Such flux enhancement dynamics demonstrate the effects of the GRB over the full energy range of the ESA (from 8 eV to ~20 keV) and over all polar and azimuthal angles (Figure 3b, c). The particle fluxes given as a function of polar and azimuthal angles in Figure 3b,c indicate that the GRB-associated secondary particles flux distributions are similar at different polar and azimuthal angles confirming the isotropy of the electron flux measurements during GRB 221009A reported by Battiston et al. (2023). The fluxes recorded during the spacecraft spin rotation combined from THB and THC SST and ESA are shown in Figure 4a and Figure 4b respectively with 0.125 second temporal resolution, which we decimate to 0.25 second resolution by accumulating 32 energy channels and data from the two spacecraft for the higher significance of the results. The distance between THB and THC corresponded to ~0.1 light second during GRB221009A and taking into account the geometry of the source (Pillera et al. 2022) the actual maximal time shift was ~0.07 s, so we combined measurements from THB and THC into a merged time series.

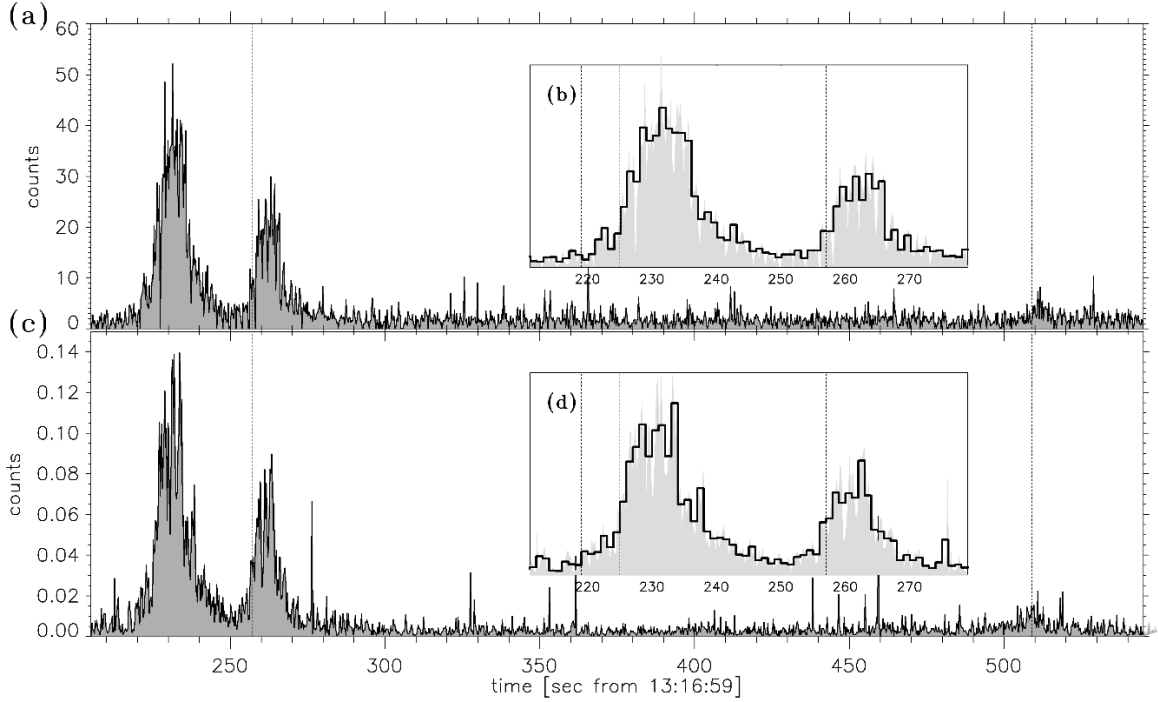

**Figure 4.** (a) - the light curve of the combined THB and THC ESA proton flux counts in the energy range from 8 eV to 20 keV. (b) – zoomed in plot of the two main bursts from (a) with curve shown at 1-second resolution. (c) - the light curve of the combined THB and THC SST proton flux counts in the energy range from 30 keV to 1 MeV. (d) – zoomed in plot of the two main bursts from (c) with solid curve shown at 1-second resolution.

The light curve of counts from THB and THC ESA detectors is presented in Figure 4a, and the combined flux light curve recorded by the SST detectors is shown in Figure 4b. The triggering impulse at 13:16:59 UT (T0 in the following) is not seen on the background of particles flux perturbation and the three peaks are resolved continuously without any signs of saturation: the shoulder beginning of the first main peak is at T0+219 s (SST) and at T0+220 s (ESA); the first main peak at T0+225 s; the second peak is at T0+256 s (SST and ESA), and the third peak is at T0+509 (SST) and T0+511 (ESA). The time marks of the GRB 221009A temporal profile reported by Battiston et al. (2023) from HEPP-L and by Liu et al. (2022) the HEBS data collected on board the LEO spacecraft (T0+219 s for the soulder, T0+225 s for the first peak, T0+256 s for the second peak, and T0+509 for third peak are indicated by the vertical dashed lines). The timing uncertainties of THEMIS spacecraft synchronization (~0.1 s) and time resolution do not allow us to resolve the time shift between the GRB signal recorded on board THB and THE and the data recorded on board the LEO spacecraft presented in detail by Battiston et al. (2023). Thus, we present a 1-second resolution version of the ESA and SST data to compare with the 1-second light curves of HEPP-L (Battiston et al. 2023) and HEBS (Liu et al. 2022). The two main peaks are zoomed in Figure 4b (ESA) and Figure 4d (SST). The light curve of 1-second resolution ESA counts in Figure 4b is in very good agreement with the 1-second resolution light curve of GRB 221009A profile presented by Liu et al. (2022) and Battiston et al. (2023): the first two sub-peaks of the first burst (T0+225-236) s are similarly resolved by ESA and SST as well as by HEPP-L and HEBS; SST resolved two later sub-peaks (the later one is seen in the HEPP-L light curve), which can be seen in the 0.05 s resolution HEBS data; the second burst at (T0+256-266 s) has 4 sub-peaks (3 from SST) confirming the measurements from HEPP-L and the 0.05 s resolution HEBS data. The good agreement with the direct observations from the gamma ray detector HEBS indicates that charged particle detectors can indeed provide highly useful observations of intense gamma bursts with , owing to their high time resolution of measurements, higher saturation thresholds, and long base lines, that are complementary to LEO spacecraft equipped with gamma ray telescopes

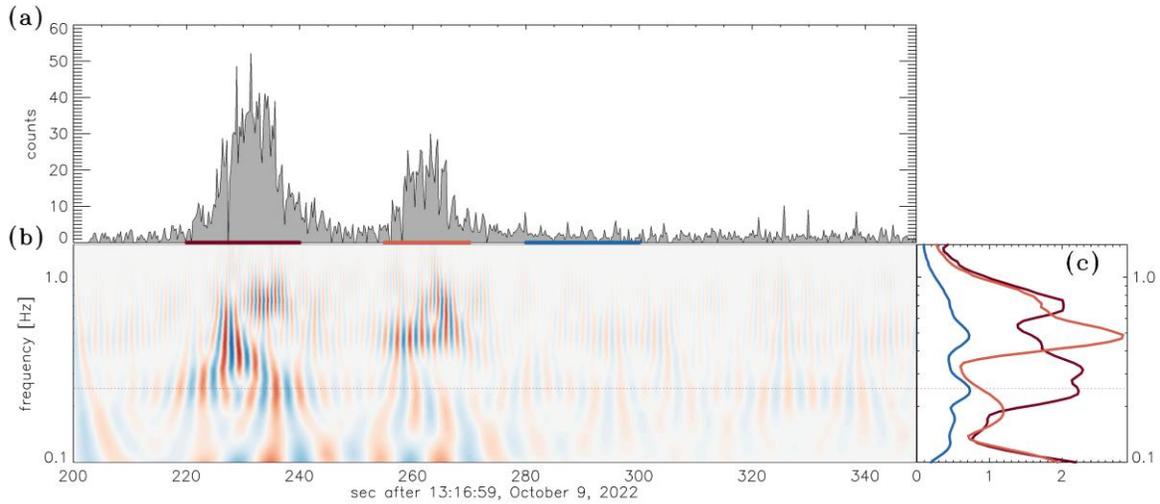

**Figure 5. (a)** – the light curve of composite THB and THC ESA proton flux counts showing the fine structure of the two main bursts of GRB221009A. (b) – wavelet decomposition of the light curve (b). The dashed line indicates the spacecraft spin frequency. (c) – time averaged wavelet spectra collected during the interval indicated by the corresponding color in panel (a).

From the 0.25 s sampling rate data, we note that the main peaks have complicated spiky fine structure. The two main bursts are zoomed in Figure 5a. The wavelet decomposition (based on the Morlet wavelet) reveals the scales of pulses indicating the pulse widths and inter-pulse time intervals - presented in Figure 5b. Figure 5d shows the time averaged wavelet spectra color-coded according to corresponding averaging intervals highlighted in Figure 5a. The first main burst has these scales to lie between 0.7 – 1.0 s, with a mean scale of $0.75_{0.63}^{0.91}$ s. The second main burst has two groups of scales with mean values of $0.5_{0.45}^{0.55}$ s and $0.75_{0.7}^{0.8}$ s (the smaller scale of ~0.5 s is close to the the double spin frequency, so its significance is not clear). These scales are in the range of the pulse scales reported by Bhat et al. (2010), where it was shown that the mean values of pulse widths were 0.81 s for long bursts and 0.04 s for short bursts, respectively: the observed GRBs statistically grouped to longer and shorter duration events with a minimum around 2 s suggesting that there are two separate populations of bursts (Bhat et al. 2010).

We have searched for the effects of GRB 2210009A in the observations of other magnetosphere and solar wind missions carrying on board charged particle detectors and we found the effects of GRB 2210009 recorded by WIND and GOES15 spacecraft. The processing of these data will be a subject of a future publication.

**Conclusions**

We present the first report of the effects of the powerful Gamma Ray Burst GRB221009A by the spacecraft particles detectors aboard the probes of the NASA THEMIS and ARTEMIS. The four spacecraft (two inner Earth's magnetosphere probes, THA and THE, and two spacecraft orbiting the Moon, THB and THC) detected the event through their electro-static analyzer (ESA) and solid state telescope (SST) proton and electron flux measurements. By combining the energy channels and the multiple spacecraft data, the fine structure of the gamma flare has been resolved with time sampling of 4 measurements per second, which make particle detectors to be additional instruments for addressing the fine structure of the intense GRBs. The obtained time scales of the fine structure spikes of the two main GRB221009A are consistent with the characteristic parameters of long gamma ray bursts.

**Acknowledgements**

We acknowledge NASA contract NAS5-02099 for use of data from the THEMIS Mission, D. Larson for use of SST data and J. P. McFadden for use of ESA data. O.V.A is grateful to J. P. McFadden for the helpful discussion. O.V.A was supported by NSF grant number 1914670, NASA's Living with a Star (LWS) program (contract 80NSSC20K0218), and NASA grants contracts 80NNSC19K0848, 80NSSC22K0433, 80NSSC22K0522, 80NSSC20K0697, and 80NSSC20K0697.